\documentclass[letterpaper]{article} 
\usepackage{aaai2027}
\usepackage[hyphens]{url}  
\usepackage{graphicx} 
\usepackage{natbib}  
\usepackage{caption} 
\usepackage[table]{xcolor}
\usepackage{times}                   
\usepackage{helvet}                  
\usepackage{courier}                 
\usepackage[hyphens]{url}            
\usepackage{graphicx}                
\usepackage{enumitem}
\usepackage{booktabs}
\usepackage{algorithm}
\usepackage{algorithmic}
\usepackage{amsmath}
\usepackage{amssymb}
\usepackage{newfloat}
\usepackage{listings}
\usepackage{natbib}                  
\usepackage{caption}                 
\usepackage{multirow}
\usepackage{makecell}
\floatstyle{ruled}
\newfloat{listing}{tb}{lst}{}
\floatname{listing}{Listing}

\DeclareCaptionStyle{ruled}{labelfont=normalfont,labelsep=colon,strut=off}
\floatstyle{ruled}
\newfloat{listing}{tb}{lst}{}
\floatname{listing}{Listing}

\title{\textsc{WhisperRec}: Latent Reasoning for Efficient Foundation Recommendation Models}

\author {
Hao Jiang\textsuperscript{\rm 1} \thanks{Equal contributions},
Peiru Du\textsuperscript{\rm 1} \footnotemark[1],
Pengfei Yao\textsuperscript{\rm 1} \footnotemark[1],
Mengting Li\textsuperscript{\rm 1}\thanks{Work done during an internship at Kuaishou Technology},
Siyuan Lou\textsuperscript{\rm 1},
Kuo Cai\textsuperscript{\rm 1},
Sheng Yu\textsuperscript{\rm 1},
Qiang Luo\textsuperscript{\rm 1}\thanks{Corresponding authors},
Jian Liang\textsuperscript{\rm 1},
Ruiming Tang\textsuperscript{\rm 1}\footnotemark[3],
Fei Pan\textsuperscript{\rm 1},
Peng Jiang\textsuperscript{\rm 1},
Wenwu Ou\textsuperscript{\rm 1}
}
\affiliations {
\textsuperscript{\rm 1}Kuaishou Technology, Beijing, China\\
\{jianghao11, dupeiru, yaopengfei03, limengting03, lousiyuan05, caikuo, yusheng03, luoqiang, liangjian03, tangruiming, panfei05, jiangpeng, luocheng10 \}@kuaishou.com
}

\vspace{-0.2cm}
\usepackage{bibentry}

\definecolor{oursgray}{RGB}{238,238,238}
\definecolor{deltagray}{RGB}{248,248,248}
\definecolor{gainred}{RGB}{150,70,55}

\begin{document}

\maketitle
\begin{abstract}
Large language models (LLMs) have demonstrated strong reasoning capabilities, motivating their adoption as the backbone of foundation recommendation models (FRMs). Existing approaches typically improve recommendation through explicit Chain-of-Thought (CoT) reasoning under the \emph{Think-then-Answer} paradigm. However, explicit CoT incurs substantial inference overhead due to lengthy reasoning generation and relies on manually designed templates that struggle to capture diverse and dynamic user interests in real-world recommendation. To address these limitations, we propose \textsc{WhisperRec}, an efficient latent reasoning framework for FRMs. The key idea is to compress explicit reasoning traces into several learnable latent reasoning tokens, enabling efficient \emph{Latent Reason-then-Answer} inference without generating verbose reasoning traces. Specifically, \textsc{WhisperRec} first introduces Multi-View Adaptive CoT (MV-ACoT), which jointly constructs diverse, high-quality teacher-generated CoT supervision by exploring user interests from multiple perspectives and automatically adapts reasoning complexity to each instance. This strategy enables the teacher model to apply lightweight reasoning to simple cases and more complex reasoning to challenging cases. Building on pre-trained FRMs, WhisperRec then employs a Three-stage Latent Token Alignment paradigm to progressively internalize teacher-generated CoT into a learnable latent token. Finally, a multi-stage curriculum-based post-training strategy effectively activates latent-token reasoning for downstream recommendation. Extensive experiments on an industrial-scale Kuaishou dataset and the public Kuaishou LLM-Rec benchmark show that \textsc{WhisperRec} consistently outperforms explicit CoT-based methods and traditional baselines. Compared with the explicit CoT \textit{Think} and \textit{No-Think} variants, \textsc{WhisperRec} improves SID@64 by 17.44\% and 9.33\%, respectively, while achieving more than $10\times$ higher online inference throughput\footnote{We will release our code upon publication.}. 
\end{abstract}


\section{Introduction}

\begin{figure}[t]
  \centering
  \includegraphics[ width=\linewidth, height=4cm]{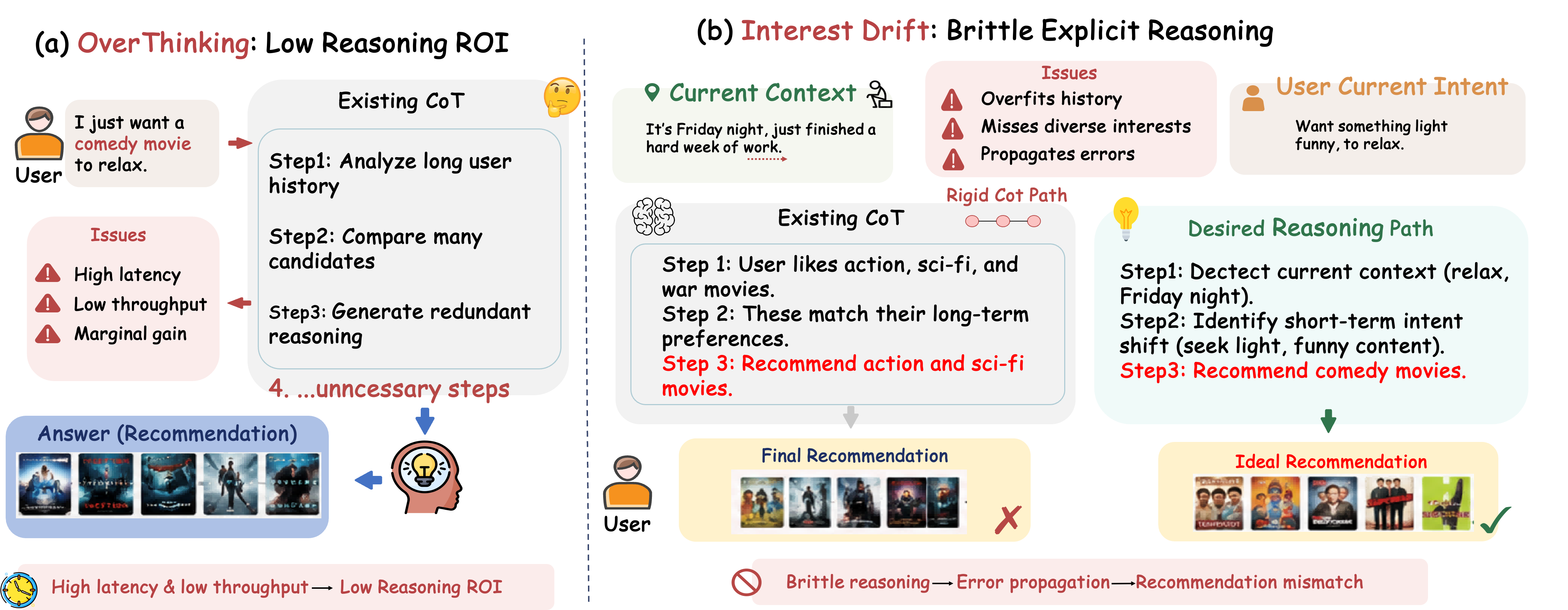}
  \vspace{-5mm}
  \caption{Limitations of explicit CoT for recommendation: overthinking of low reasoning ROI and interest drift.}
  \label{intro}
  \vspace{-0.5cm}
\end{figure}

Driven by the scaling of model capacity and data, LLMs have progressed from generation to knowledge-intensive reasoning and agentic decision making~\cite{jaech2024openai, guo2025deepseek}. This progress has motivated researchers to transfer and inherit these capabilities to recommendation systems~\cite{zhai2024actions, zhang2024wukong, rajput2018recommender}. Recent generative recommendation models, exemplified by the \textsc{OneRec} family~\cite{zhou2025onerec, zhou2025onerecv2, liu2025onerecthink, zhou2025openonerec}, have demonstrated that Transformer-based architectures can scale recommendation modeling to support industrial-scale end-to-end deployment. However, they mostly make predictions through implicit representation computing, making their decision processes difficult to interpret or control~\cite{zhou2017deep, zhou2019deep}. Although several studies introduce latent reasoning mechanisms into recommendation~\cite{tang2026think, dai2025onepiece}, their reasoning still operates in a non-semantic latent space, making it difficult to determine whether the learned representations encode coherent, task-relevant reasoning.

A direct approach is to directly use LLMs as the backbone of foundation recommendation models (FRMs) and adapt them through recommendation-specific training~\cite{yi2025recgpt, yi2025recgptv2, zhou2025openonerec}. Recent work, particularly \textsc{OneReason}~\cite{team2026onereason}, demonstrates the sucess of the \emph{Think-then-Answer} paradigm for FRMs, in which the model reasons over user preferences and context before generating recommendations. However, recommendation reasoning cannot simply inherit generic CoT patterns~\cite{zhou2025perception, zhang2026thinking}, as user interests are diverse and evolving, requiring structured reasoning from preference induction to intent inference and decision making. Despite its promise, explicit CoT-based FRMs face two key limitations.

\textbf{(1) Overthinking and low reasoning ROI.}
As shown in Figure~\ref{intro}(a), explicit CoT often generates lengthy and redundant reasoning even for simple requests with clear intents. This increases inference latency and reduces throughput, while bringing only marginal recommendation gains.

\textbf{(2) Interest drift and brittle explicit reasoning.}
As shown in Figure~\ref{intro}(b), fixed and single-path of CoT can overemphasize dominant historical preferences while overlooking short-term, context-dependent interests. Once even a minor reasoning step goes wrong, the error may propagate to the final recommendation and cause a mismatch.

To address these limitations, we propose \textsc{WhisperRec}, a latent reasoning framework for FRMs. Rather than laboriously generating explicit CoT for each recommendation decision, \textsc{WhisperRec} internalizes recommendation reasoning capabilities into a set of learnable latent tokens, enabling a \emph{Latent-Reason-then-Answer} paradigm. This design preserves high-dimensional latent reasoning capacity while avoiding the latency and throughput bottleneck of autoregressive rationale generation. Specifically, \textsc{WhisperRec} first introduces Multi-View Adaptive CoT (MV-ACoT) to construct teacher-generated CoT supervision. MV-ACoT jointly explores user interests from multiple perspectives and automatically adapts reasoning complexity to individual instances, applying lightweight reasoning to simple cases and targeted reasoning to challenging ones. Building on a pre-trained FRM, \textsc{WhisperRec} then employs a Three-stage Latent Token Alignment paradigm to progressively distill teacher-generated CoT into learnable latent tokens. Finally, a multi-stage curriculum-based post-training strategy activates latent-token reasoning for downstream recommendation, enabling efficient and robust recommendations under dynamic and complex user interest dynamics.

We conduct extensive experiments on an industrial-scale Kuaishou recommendation dataset and the public Kuaishou LLM-Rec benchmark. \textsc{WhisperRec} consistently outperforms competitive recommendation baselines, including the strong explicit CoT-based \textsc{OneReason}. In particular, it improves SID@64 by 17.44\% over the explicit Think variant and by 9.33\% over the No-Think variant, while delivering more than $10\times$ higher online inference throughput than explicit reasoning. Further analyses show that MV-ACoT improves teacher-generated CoT quality, leading to significant gains in recommendation performance. Latent Token Alignment transfers explicit CoT knowledge into latent tokens, while multi-stage curriculum-based post-training further activates latent reasoning. We also observe a strong positive correlation between CoT quality and  recommendation performance, suggesting that reasoning benefits recommendation by improving decision-relevant representations rather than by generating lengthy natural-language rationales.

Our main contributions are summarized as follows:
\begin{itemize}

\item We propose \textsc{WhisperRec}, an efficient latent reasoning framework that introduces the \emph{Latent Reason-then-Answer} paradigm for FRMs and improves performance.

\item We develop MV-ACoT, a Latent Reasoning Alignment framework, and curriculum-based post-training to internalize explicit CoT into learnable latent tokens for FRMs.

\item Extensive experiments on industrial and public benchmarks show that \textsc{WhisperRec} achieves superior recommendation performance and approximately $10\times$ higher inference throughput than explicit CoT methods.

\end{itemize}

\section{Related Work}

\subsection{Scaling Recommendation with LLMs}
Traditional recommendation methods learn user representations from historical interactions and predict the next item~\citep{hidasi2015session, kang2018self, sun2019bert4rec}. With the rise of LLMs, early LLM-based approaches directly prompted LLMs to make recommendation decisions~\citep{geng2022recommendation, gao2023chat, hou2024large}. Although these methods leverage the world knowledge of LLMs, they remain limited in modeling collaborative signals.
Recent studies have advanced recommendation systems toward foundation recommendation models along two complementary directions. The first develops recommendation-native backbones: TIGER represents items using semantic IDs~\citep{rajput2023recommender}, HSTU formulates large-scale recommendation as sequence transduction~\citep{zhai2024actions}, and OneRec unifies retrieval and ranking in an end-to-end generative architecture~\citep{deng2025onerec}. The second adapts pretrained LLMs to recommendation data. LC-Rec aligns linguistic knowledge with collaborative semantics~\citep{zheng2024adapting}, while the RecGPT family develops transferable LLM-based recommenders through recommendation-domain pre-training~\citep{yi2025recgpt, yi2025recgptv2}. Despite their success, these models mainly focus on behavior modeling and item generation, providing limited supervision for inferring latent user needs and making recommendation decisions.

\subsection{Reasoning-Enhanced Recommendation}
Recent work has begun to equip recommendation models with reasoning capabilities. Latent reasoning methods reduce inference costs by avoiding lengthy natural-language rationales. ReaRec recursively refines user representations in the hidden space~\citep{tang2026think}, OnePiece introduces reasoning modules into industrial ranking models~\citep{dai2025onepiece}, OneSearch-V2 distills LLM reasoning into compact internal states~\citep{chen2026onesearch}, and REG4Rec explores multiple latent reasoning paths through self-reflection~\citep{xing2025reg4rec}.
In parallel, explicit reasoning methods expose intermediate reasoning processes in natural language. GOT4Rec~\citep{long2024got4rec}, Reason4Rec~\citep{fang2025reason4rec}, and ThinkRec~\citep{yu2026thinkrec} construct user preference analyses or multi-path reasoning trajectories through supervised fine-tuning. RecOne~\citep{kong2026think}, OneRec-Think~\citep{liu2025onerec}, and OpenOneRec~\citep{zhou2025openonerec} further combine CoT supervision with reinforcement learning to improve recommendation decisions. OneReason~\citep{team2026onereason} shows that simply introducing explicit CoT does not consistently activate reasoning in FRMs, and strengthens the reasoning process through item semantic awareness, user behavior cognition, and reinforcement learning.
Although these methods advance reasoning-based recommendation, explicit reasoning is costly and prone to noisy rationales, whereas latent reasoning often lacks direct supervision for recommendation decisions. Our work bridges both paradigms by internalizing diverse, domain-specific reasoning trajectories into latent tokens through pre-training and post-training.

\section{Methodology}

\begin{figure*}[tp]
  \centering
  \vspace{-0.3cm}

  \includegraphics[width=\textwidth,     height=6.5cm
]{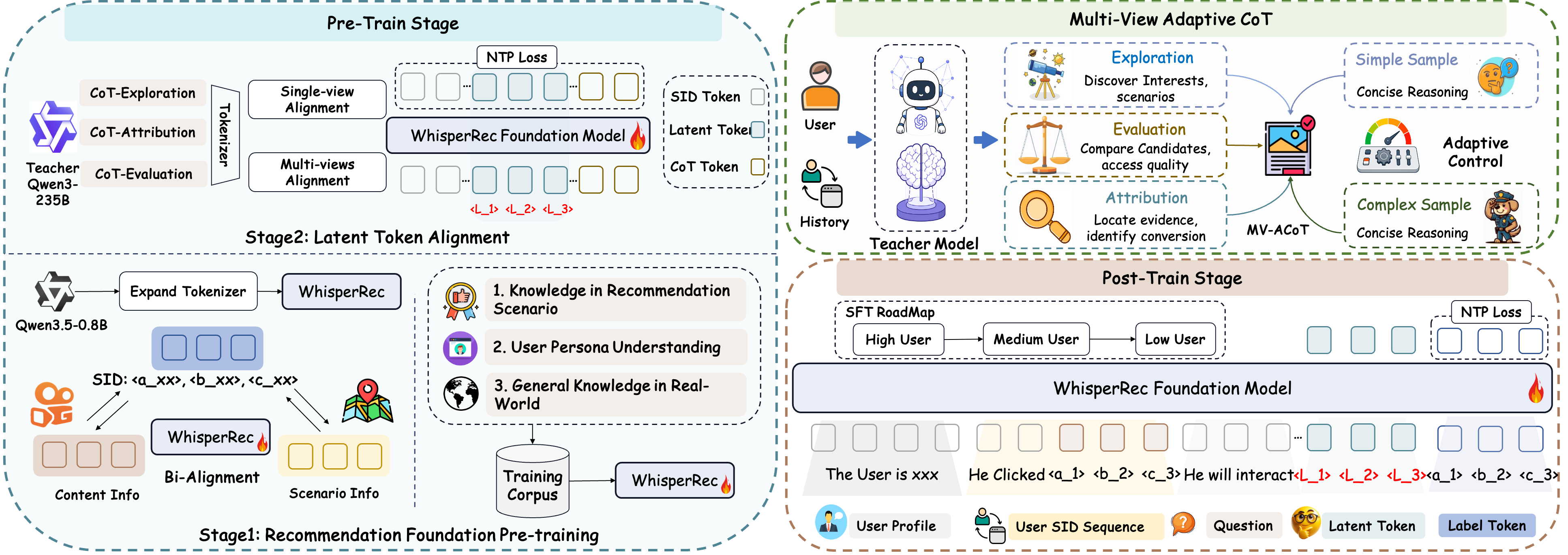}
  \caption{Overview of the \textsc{WhisperRec}, containing Multi-view Adaptive CoT, Latent Token Alignment and Post-training stage.}
  \label{fig:framework}
  \vspace{-0.5cm}
\end{figure*}

\subsection{Problem Formulation and Framework}

For a user $u$, let $P_u$ denote the user profile, $H_u$ the chronologically ordered interaction history, $\mathcal{C}_u$ the optional spatiotemporal context, and $i_u^+$ the next interacted item. Each item $i$ is represented by a Semantic ID (SID) sequence $\mathcal{S}_i=[s_i^{(1)};s_i^{(2)};s_i^{(3)}]$. The recommendation objective is to generate $\mathcal{S}_{i_u^+}$ from information available before the target interaction. Neither $i_u^+$ nor its SID is included in the  input.

As shown in Figure~\ref{fig:framework}, \textsc{WhisperRec} comprises three components. First, Multi-View Adaptive Chain-of-Thought constructs complementary teacher-generated supervision for intent exploration, candidate evaluation, and interaction attribution. Second, Latent Token Alignment distills these rationales into a small set of contextualized latent tokens. Third, curriculum-based post-training jointly optimizes standard recommendation and latent-reasoning modes. During online inference, the model takes the user context and latent tokens as input and directly generates the target SID, avoiding autoregressive natural-language rationale generation.

\subsection{Multi-View Adaptive CoT}
\label{sec:Multi-View Adaptive CoT}
In this section, we propose MV-ACoT to construct high-quality, recommendation-oriented CoT supervision that equips FRMs with latent reasoning capabilities.

\subsubsection{Multi-View CoT Task Design}
We argure that effective recommendation CoT should capture decision-relevant signals without unnecessary complexity. However, under multi-interest, context-dependent, and noisy user behaviors, a single reasoning task may lead to uncontrolled divergence, inaccurate candidate assessment, or hindsight rationalization. MV-ACoT addresses this issue by constructing complementary \textbf{Exploration}, \textbf{Evaluation}, and \textbf{Attribution} traces, while adapting reasoning complexity to sample-level decision difficulty to avoid overthinking on easy cases and insufficient analysis on hard ones.

Given a user profile $U$, historical behavior sequence $H$, candidate ad item $i_c$, difficulty of instance $d$, ad item $i_c$ and target item $i^+$, the teacher model $P_{\theta}$ generates reasoning traces $T$ and auxiliary judgment signals under different CoT prompt functions.

\paragraph{View 1: Exploration}
The exploration view is conditioned only on the user profile and item interaction history, without access to a candidate or target item, and infers plausible commercial intents:
\begin{equation}
(R_u^{\mathrm{exp}},\mathcal{I}_u)
\sim
p_{\mathrm{T}}\!\left(
\cdot \mid \pi_{\mathrm{exp}}(P_u,H_u;d_u)
\right),
\label{eq:exploration}
\end{equation}
where $R_u^{\mathrm{exp}}$ is the rationale, $\mathcal{I}_u$ is a set of potential intents. This strategy encourages model to infer diverse user needs, interest shifts, and exploratory recommendation directions, thereby expanding the latent intent space.

\paragraph{View 2: Evaluation}
Given a candidate $i_c$, the evaluation view assesses whether it is supported by the user's interests:
\begin{equation}
(R_{u,c}^{\mathrm{eval}},y_{u,c},q_{u,c})
\sim
p_{\mathrm{T}}\!\left(
\cdot \mid
\pi_{\mathrm{eval}}(P_u,H_u,\mathcal{C}_u,i_c;d_{u,c})
\right),
\label{eq:evaluation}
\end{equation}
where $y_{u,c}$ is a categorical matching judgment and $q_{u,c}\in[0,1]$ is the teacher confidence. This strategy simulates user-item pair evaluation in recommendation ranking and allows negative judgments when the evidence is insufficient.

\paragraph{View 3: Attribution}
For an observed positive interaction $i_u^+$, the attribution view identifies the preceding evidence that supports this outcome:
\begin{equation}
(R_u^{\mathrm{attr}},a_u,q_u^{\mathrm{attr}})
\sim
p_{\mathrm{T}}\!\left(
\cdot \mid
\pi_{\mathrm{attr}}(P_u,H_u,\mathcal{C}_u,i_u^+;d_u)
\right),
\label{eq:attribution}
\end{equation}
where $a_u$ denotes the attribution category and $q_u^{\mathrm{attr}}\in[0,1]$ is the attribution confidence. This strategy anchors reasoning on the observed conversion and captures evidence chains behind realized commercial intent.

The three strategies provide complementary supervision. Exploration supports open-ended intent discovery, evaluation enables candidate discrimination without assuming correctness, and attribution grounds reasoning in conversion evidence while mitigating hindsight rationalization. Together, they provide generative, discriminative, and evidence-based CoT supervision, reducing the bias of a single reasoning task.

\subsubsection{Adaptive CoT Generation}

Beyond multi-view construction, MV-ACoT adapts reasoning complexity to sample-level decision difficulty. Unlike fixed-template CoT, which applies the same reasoning process to all samples, MV-ACoT estimates the complexity of a user context from behavioral signal strength, weakly related behaviors, and noisy interactions:
\begin{equation}
d = g(U, H), \quad d \in \{\textsc{Low}, \textsc{High}\},
\end{equation}
where $g(\cdot)$ denotes the complexity assessment function.

Given the difficulty $d$, the teacher model generates a reasoning trace conditioned on an adaptive prompt $\pi_d$:
\begin{equation}
T \sim P_{\theta}\left(
T \mid U, H, i_c, i^+; \pi_d, \mathcal{C}
\right),
\end{equation}
where $\mathcal{C}$ denotes recommendation constraints, including behavioral strength, fulfillment mode, spatiotemporal context, and industry-specific factors. For high-uncertainty samples, $\pi_d$ expands reasoning across multiple factors to distinguish competing intents and identify noisy signals. For low-uncertainty samples, it prunes unnecessary dimensions and focuses on dominant conversion signals. Thus, MV-ACoT allocates its reasoning budget according to decision difficulty rather than uniformly generating longer CoT. Appendix C summarizes the prompt construction principles.

\subsection{Latent Token Alignment}
Starting from the pre-trained FRMs $\mathcal{F}_{\mathrm{base}}$, we conduct a three-stage Latent Reasoning Alignment procedure, containing single-path and multi-path latent reasoning alignment and recommendation-oriented context alignment. The first two stages distill teacher-generated CoT into latent tokens under teacher model supervision, while the final stage aligns the learned latent token with next-item prediction.

\subsubsection{FRMs Pre-training}
We initialize from Qwen3.5-0.8B and extend its vocabulary with Semantic ID tokens. Each item $v$ is represented by a three-level SID sequence with Res-Kmeans:
\begin{equation}
\mathcal{S}_v =
[s_v^{(1)}; s_v^{(2)}; s_v^{(3)}],
\end{equation}
where $s_v^{(l)}$ denotes the SID at level $l$, encoding item side multi-model information~\cite{luo2025qarm}. We bidirectionally align $\mathcal{S}_v$ with the dense item's caption $C_v$:
\begin{equation}
\mathcal{L}_{\mathrm{sem}}
=
-\log p_{\theta}(\mathcal{S}_v \mid C_v)
-\log p_{\theta}(C_v \mid \mathcal{S}_v).
\end{equation}
Moreover, we introduce diverse CPT tasks to enhance the FRMs, including recommendation-domain knowledge such as spatiotemporal context, user persona understanding, and general-domain knowledge. Together, these tasks construct a capable foundation model for recommendation.

\subsubsection{Stage I: Single-View Latent Reasoning Alignment}

We first warm up the latent tokens using one high-confidence, target-free rationale for each training context. Let $v_u^\star$ denote the selected reasoning view and $\langle v_u^\star\rangle$ its corresponding control token. The training sequence is constructed as
\begin{equation}
[X_u;\mathcal{L};\langle v_u^\star\rangle;
\widetilde{R}_u^{(v_u^\star)}].
\end{equation}
We apply the training loss only to the rationale tokens:
\begin{equation}
\mathcal{L}_{\mathrm{single}}
=
-\mathbb{E}_{u}
\sum_{t=1}^{|\widetilde{R}_u^{(v_u^\star)}|}
\log p_\phi\!\left(
\widetilde{r}_{u,t}^{(v_u^\star)}
\mid
X_u,\mathcal{L},\langle v_u^\star\rangle,
\widetilde{r}_{u,<t}^{(v_u^\star)}
\right).
\label{eq:single}
\end{equation}
Placing the view token after the shared latent tokens ensures that the latent slots are derived from a common recommendation context, rather than being conditioned on view-specific textual instructions.

\subsubsection{Stage II: Multi-Views Latent Reasoning Alignment}

We then use every retained MV-ACoT rationale. Let
$\mathcal{V}_u\subseteq
\{\mathrm{exp},\mathrm{eval},\mathrm{attr}\}$ be the available views for user
$u$. The multi-view objective is
Let
$C_{u,t}^{(v)}
=(X_u,\mathcal{L},v,\widetilde{r}_{u,<t}^{(v)})$
denote the decoding context at step $t$. The multi-view objective is
\begin{equation}
\begin{aligned}
\ell_u^{(v)}
&=
-\frac{1}{|\widetilde{R}_u^{(v)}|}
\sum_{t=1}^{|\widetilde{R}_u^{(v)}|}
\log p_\phi\!\left(
\widetilde{r}_{u,t}^{(v)}
\mid C_{u,t}^{(v)}
\right),\\
\mathcal{L}_{\mathrm{multi}}
&=
\mathbb{E}_{u}\!\left[
\frac{1}{|\mathcal{V}_u|}
\sum_{v\in\mathcal{V}_u}
\ell_u^{(v)}
\right].
\end{aligned}
\label{eq:multi}
\end{equation}
Each rationale is paired with a view token indicating exploration, evaluation, or attribution. We average token losses within each rationale so that long traces do not receive greater weight. Training all three views with the same contextualized slots encourages a shared user representation that supports all three reasoning objectives.

\subsubsection{Stage III: Recommendation-Oriented Context Alignment}

The preceding stages teach latent tokens to represent CoT knowledge, but do not directly optimize recommendation decisions. We therefore align latent token with target-item prediction. In addition to the user profile $P_u$ and historical SID sequence $H_u$, we introduce recommendation contextual information $\mathcal{C}_u$ (e.g., temporal and geographical signals). The target SID sequence is excluded from the input:
\begin{equation}
X_u^{\mathrm{ctx}} =
[\mathcal{Z}; P_u; H_u; \mathcal{C}_u].
\end{equation}
The model predicts the observed target SID sequence $\mathcal{S}_u^+$:
\begin{equation}
\mathcal{L}_{\mathrm{align}}
=
-\mathbb{E}
\left[
\log p_{\theta}
\left(
\mathcal{S}_u^+ \mid X_u^{\mathrm{ctx}}
\right)
\right].
\end{equation}
This stage makes latent tokens a bridge between user context and recommendation decisions, yielding the context-aware latent reasoning FRM $\mathcal{F}_{\mathrm{latent\text{-}ctx}}$.

\begin{table*}[tp]
\centering

\begingroup
\tiny
\setlength{\tabcolsep}{1.5pt}
\renewcommand{\arraystretch}{0.80}
\resizebox{\textwidth}{!}{
\begin{tabular}{lllcccccccccccccccc}
\toprule
\multirow{2}{*}{\textbf{Backbone}}
& \multirow{2}{*}{\textbf{Method}}
& \multirow{2}{*}{\textbf{CoT}}
& \multicolumn{8}{c}{\textbf{OneReason Public Benchmark}}
& \multicolumn{8}{c}{\textbf{Industrial Dataset}} \\
\cmidrule(lr){4-11} \cmidrule(lr){12-19}
& & 
& \textbf{SID@1} & \textbf{SID@16} & \textbf{SID@32} & \textbf{SID@64}
& \textbf{ID@1} & \textbf{ID@16} & \textbf{ID@32} & \textbf{ID@64}
& \textbf{SID@1} & \textbf{SID@16} & \textbf{SID@32} & \textbf{SID@64}
& \textbf{ID@1} & \textbf{ID@16} & \textbf{ID@32} & \textbf{ID@64} \\
\midrule

\multirow{3}{*}{ID-based}
& GRU4Rec
& / & -- & -- & -- & -- & 0.0119 & 0.0147 & 0.0157 & 0.0182
& -- & -- & -- & -- & 0.0158 & 0.1057 & 0.1454 & 0.1883 \\

& BERT4Rec
& / & -- & -- & -- & -- & 0.0122 & 0.0153 & 0.0161 & 0.0189
& -- & -- & -- & -- & 0.0174 & 0.1074 & 0.1480 & 0.1932 \\

& HSTU
& / & -- & -- & -- & -- & 0.0114 & 0.0146 & 0.0165 & 0.0195
& -- & -- & -- & -- & 0.0145 & 0.1039 & 0.1414 & 0.1812 \\
\midrule

\multirow{2}{*}{SID-based}
& TIGER
& / & 0.0051 & 0.0061 & 0.0076 & 0.0078 & 0.0106 & 0.0124 & 0.0137 & 0.0156
& 0.0714 & 0.1000 & 0.1197 & 0.1432 & 0.0092 & 0.0401 & 0.0598 & 0.0859 \\

& OneRec
& / & 0.0049 & 0.0063 & 0.0076 & 0.0082 & 0.0108 & 0.0135 & 0.0152 & 0.0177
& 0.0708 & 0.1031 & 0.1252 & 0.1468 & 0.0116 & 0.0447 & 0.0650 & 0.0900 \\
\midrule

\multirow{1}{*}{LLM-based}
& OneReason
& / & 0.0055 & 0.0077 & 0.0079 & 0.0081 & 0.0234 & 0.0244 & 0.0258 & 0.0272
& 0.0762 & 0.1527 & 0.1603 & 0.1658 & 0.2254 & 0.4522 & 0.4750 & 0.4941 \\
\midrule

\multirow{6}{*}{LLM-based}
& \multirow{6}{*}{\makecell[c]{OneReason+CoT\\NoThink}}
& OR
& 0.0065 & 0.0103 & 0.0104 & 0.0104 & 0.0262 & 0.0306 & 0.0314 & 0.0334
& 0.0809 & 0.1639 & 0.1739 & 0.1823 & 0.2293 & 0.4499 & 0.4727 & 0.4936 \\

& & Ada
& 0.0070 & 0.0113 & 0.0124 & 0.0130 & 0.0270 & 0.0341 & 0.0364 & 0.0387
& 0.0873 & 0.1746 & 0.1832 & 0.1913 & 0.2365 & 0.4620 & 0.4850 & 0.5049 \\

& & Exp.
& 0.0068 & 0.0112 & 0.0113 & 0.0113 & 0.0228 & 0.0286 & 0.0318 & 0.0374
& 0.0881 & 0.1728 & 0.1819 & 0.1894 & 0.2385 & 0.4621 & 0.4845 & 0.5032 \\

& & Eval.
& 0.0079 & 0.0139 & 0.0172 & 0.0213 & 0.0252 & 0.0357 & 0.0486 & 0.0623
& 0.0894 & 0.1800 & 0.1896 & 0.1962 & 0.2356 & 0.4630 & 0.4881 & 0.5065 \\

& & Attr.
& 0.0071 & 0.0125 & 0.0165 & 0.0208 & 0.0255 & 0.0379 & 0.0462 & 0.0601
& 0.0897 & 0.1793 & 0.1876 & 0.1943 & 0.2392 & 0.4640 & 0.4860 & 0.5040 \\

& & Merge
& 0.0082 & 0.0131 & 0.0178 & 0.0209 & 0.0258 & 0.0362 & 0.0471 & 0.0612
& 0.0861 & 0.1720 & 0.1790 & 0.1851 & 0.2385 & 0.4640 & 0.4846 & 0.5038 \\
\midrule

\multirow{6}{*}{LLM-based}
& \multirow{6}{*}{\makecell[c]{OneReason+CoT\\Think}}
& OR
& 0.0063 & 0.0073 & 0.0080 & 0.0089 & 0.0235 & 0.0263 & 0.0278 & 0.0305
& 0.0601 & 0.1405 & 0.1563 & 0.1697 & 0.1838 & 0.4080 & 0.4450 & 0.4759 \\

& & Ada
& 0.0064 & 0.0074 & 0.0086 & 0.0093 & 0.0233 & 0.0267 & 0.0296 & 0.0311
& 0.0714 & 0.1486 & 0.1590 & 0.1688 & 0.2077 & 0.4329 & 0.4606 & 0.4844 \\

& & Exp.
& 0.0065 & 0.0082 & 0.0084 & 0.0087 & 0.0235 & 0.0285 & 0.0291 & 0.0299
& 0.0672 & 0.1190 & 0.1391 & 0.1465 & 0.2013 & 0.3992 & 0.4301 & 0.4512 \\

& & Eval.
& 0.0065 & 0.0081 & 0.0087 & 0.0096 & 0.0234 & 0.0276 & 0.0299 & 0.0328
& 0.0696 & 0.1169 & 0.1332 & 0.1487 & 0.2083 & 0.3575 & 0.3952 & 0.4303 \\

& & Attr.
& 0.0066 & 0.0078 & 0.0081 & 0.0084 & 0.0235 & 0.0273 & 0.0282 & 0.0296
& 0.0682 & 0.1287 & 0.1452 & 0.1617 & 0.2029 & 0.3739 & 0.4088 & 0.4458 \\

& & Merge
& 0.0072 & 0.0084 & 0.0086 & 0.0091 & 0.0241 & 0.0281 & 0.0288 & 0.0303
& 0.0728 & 0.1301 & 0.1400 & 0.1505 & 0.2127 & 0.3985 & 0.4275 & 0.4567 \\
\midrule

\multirow{7}{*}{LLM-based}
& \multirow{7}{*}{\textsc{WhisperRec}}
& OR
& 0.0065 & 0.0082 & 0.0089 & 0.0092 & 0.0156 & 0.0253 & 0.0268 & 0.0282
& 0.0882 & 0.1771 & 0.1867 & 0.1939 & 0.2354 & 0.4632 & 0.4856 & 0.5061 \\

& & Ada
& 0.0075 & 0.0126 & 0.0138 & 0.0144 & 0.0224 & 0.0378 & 0.0403 & 0.0429
& 0.0898 & 0.1770 & 0.1869 & 0.1935 & 0.2376 & 0.4631 & 0.4867 & 0.5052 \\

& & Exp.
& 0.0072 & 0.0125 & 0.0137 & 0.0144 & 0.0179 & 0.0350 & 0.0373 & 0.0402
& 0.0930 & 0.1811 & 0.1908 & 0.1971 & 0.2425 & 0.4650 & 0.4883 & 0.5060 \\

& & Eval.
& 0.0087 & 0.0155 & 0.0189 & 0.0233 & 0.0300 & 0.0410 & 0.0560 & 0.0712
& 0.0946 & 0.1831 & 0.1931 & 0.1989 & 0.2436 & 0.4660 & 0.4891 & 0.5070 \\

& & Attr.
& 0.0072 & 0.0145 & 0.0150 & 0.0167 & 0.0299 & 0.0365 & 0.0478 & 0.0560
& 0.0902 & 0.1779 & 0.1869 & 0.1936 & 0.2396 & 0.4645 & 0.4879 & 0.5055 \\

& & \cellcolor{oursgray}\textbf{Merge}
& \cellcolor{oursgray}\textbf{0.0089}
& \cellcolor{oursgray}\textbf{0.0158}
& \cellcolor{oursgray}\textbf{0.0193}
& \cellcolor{oursgray}\textbf{0.0239}
& \cellcolor{oursgray}\textbf{0.0331}
& \cellcolor{oursgray}\textbf{0.0506}
& \cellcolor{oursgray}\textbf{0.0584}
& \cellcolor{oursgray}\textbf{0.0742}
& \cellcolor{oursgray}\textbf{0.0949}
& \cellcolor{oursgray}\textbf{0.1847}
& \cellcolor{oursgray}\textbf{0.1957}
& \cellcolor{oursgray}\textbf{0.1993}
& \cellcolor{oursgray}\textbf{0.2440}
& \cellcolor{oursgray}\textbf{0.4662}
& \cellcolor{oursgray}\textbf{0.4895}
& \cellcolor{oursgray}\textbf{0.5088} \\

& & \makecell[c]{\textit{vs. SOTA}\\\textit{($\Delta$\%)}}
& \textcolor{gainred}{\textbf{+41\%$\uparrow$}}
& \textcolor{gainred}{\textbf{+116\%$\uparrow$}}
& \textcolor{gainred}{\textbf{+141\%$\uparrow$}}
& \textcolor{gainred}{\textbf{+169\%$\uparrow$}}
& \textcolor{gainred}{\textbf{+41\%$\uparrow$}}
& \textcolor{gainred}{\textbf{+92\%$\uparrow$}}
& \textcolor{gainred}{\textbf{+110\%$\uparrow$}}
& \textcolor{gainred}{\textbf{+143\%$\uparrow$}}
& \textcolor{gainred}{\textbf{+58\%$\uparrow$}}
& \textcolor{gainred}{\textbf{+31\%$\uparrow$}}
& \textcolor{gainred}{\textbf{+25\%$\uparrow$}}
& \textcolor{gainred}{\textbf{+17\%$\uparrow$}}
& \textcolor{gainred}{\textbf{+33\%$\uparrow$}}
& \textcolor{gainred}{\textbf{+14\%$\uparrow$}}
& \textcolor{gainred}{\textbf{+10\%$\uparrow$}}
& \textcolor{gainred}{\textbf{+7\%$\uparrow$}} \\
\bottomrule
\end{tabular}
}
\endgroup

\caption{Results on public and industrial datasets. ID-based baselines do not generate semantic IDs, and their SID results are marked as ``--''. OR/Ada/Exp./Eval./Attr. denote OneReason-CoT, Adaptive CoT, and MV-ACoT views.}

\label{tab:WhisperRec_benchmark_compact}
\vspace{-0.5cm}
\end{table*}

\subsection{Post-Training and Inference}

We further adapt $\mathcal{F}_{\mathrm{latent\text{-}ctx}}$ to the target recommendation scenario through multi-stage curriculum-based recommendation SFT and latent reasoning SFT.

\subsubsection{Multi-stage Curriculum-based Recommendation SFT}
The primary objective of this stage is to equip \textsc{WhisperRec} with recommendation capabilities in the target scenario.
We first train the model to predict the next target SID sequence from standard recommendation inputs:
\begin{equation}
X_u^{\mathrm{std}} = [P_u; H_u].
\end{equation}
Here, $P_u$ provides persistent user attributes, while $H_u$ captures behavioral preferences. The training objective is
\begin{equation}
\mathcal{L}_{\mathrm{SFT}}
=
-\mathbb{E}
\left[
\log p_{\theta}
\left(
\mathcal{S}_u^+ \mid X_u^{\mathrm{std}}
\right)
\right].
\end{equation}
To smoothly equips a LLMs with recommendation capability across industrial scenarios with varying degrees of user activity sparsity, we introduce a \textbf{multi-stage curriculum learning strategy}. Specially, we employ a curriculum from high-activity to low-activity users:
\begin{equation}
\mathcal{D}_{\mathrm{high}}
\rightarrow
\mathcal{D}_{\mathrm{medium}}
\rightarrow
\mathcal{D}_{\mathrm{low}}.
\end{equation}

\subsubsection{Latent Reasoning-based Recommendation SFT}

We then activate latent reasoning by prepending the learned latent tokens to the same user context:
\begin{equation}
X_u^{\mathrm{LR}} =
[\mathcal{Z}; P_u; H_u].
\end{equation}
Unlike context alignment, this stage omits auxiliary context signals to ensure that latent reasoning remains effective under standard online recommendation inputs. The model is optimized to predict the same target SID sequence:
\begin{equation}
\mathcal{L}_{\mathrm{LR}}
=
-\mathbb{E}
\left[
\log p_{\theta}
\left(
\mathcal{S}_u^+ \mid X_u^{\mathrm{LR}}
\right)
\right].
\end{equation}

To preserve performance in both latent-reasoning and standard recommendation modes, we mix latent reasoning and conventional SFT samples at a $1{:}1$ ratio:
\begin{equation}
\mathcal{L}_{\mathrm{post}}
=
\frac{1}{2}\mathcal{L}_{\mathrm{LR}}
+
\frac{1}{2}\mathcal{L}_{\mathrm{SFT}}.
\end{equation}
This objective enables the \emph{Latent-Reason-then-Answer} paradigm when $\mathcal{Z}$ is provided, while retaining stable recommendation performance without latent tokens.

\subsubsection{Inference}

Online inference uses $X_u^{\mathrm{LR}}$ and decodes only the three-level target SID:
\begin{equation}
\widehat{\mathcal{S}}_u
=
\operatorname*{arg\,max}_{\mathcal{S}}
p_\phi(\mathcal{S}\mid P_u,H_u,\mathcal{L}).
\label{eq:inference}
\end{equation}
The MV-ACoT teacher, view tokens, privileged item information, and textual rationale decoder are absent at inference. The reasoning overhead is bounded by the fixed number $K$ of latent tokens rather than by the length of a generated natural-language CoT traces.

\section{Experiment}

We organize the experiments around four questions:
\begin{itemize}
    \item \textbf{RQ1:} Does \textsc{WhisperRec} outperform ID-based and SID-based baselines, as well as state-of-the-art FRMs?

    \item \textbf{RQ2:} Does MV-ACoT provide higher-quality and complementary CoT supervision for improving FRMs?

    \item \textbf{RQ3:} Can latent reasoning further improve performance over explicit CoT reasoning?

    \item \textbf{RQ4:} Does latent reasoning preserve general ability while improving reasoning efficiency?
\end{itemize}

\subsection{Experiment Setting}
\subsubsection{Datasets and Models}
We evaluate \textsc{WhisperRec} on two datasets. The public dataset is derived from the advertising domain of the Kuaishou LLM-Rec benchmark~\footnote{\tiny\url{https://huggingface.co/datasets/OpenOneRec/Explorer_LLM_Rec_Competition}}. The industrial dataset is collected from Kuaishou local-service scenarios and contains seven days of real-world online user interaction data. The basic statistics of the two datasets are shown in Table \ref{tab:dataset_statistics}.

We use Qwen3.5-0.8B~\cite{qwen3.5} as the FRM's backbone and extend its vocabulary with SID tokens and latent tokens. For the public benchmark, we initialize WhisperRec from the OneReason-0.8B-pretrain-competition checkpoint~\footnote{\tiny\url{https://huggingface.co/OpenOneRec/OneReason-0.8B-pretrain-competition}}~\cite{onereason2026}, followed by latent reasoning pre-training and post-training.

\begin{table}[h]
\centering
\scriptsize
\renewcommand{\arraystretch}{1.08}
\begin{tabular*}{\columnwidth}{@{\extracolsep{\fill}}lccc@{}}
\toprule
\textbf{Dataset} & \textbf{SFT Train} & \textbf{CoT Train} & \textbf{Test} \\
\midrule
Kuaishou Locallife & 100,063 & 59,113 & 20,393 \\
Kuaishou LLM-Rec & 60,000 & 30,000 & 10,000 \\
\bottomrule
\end{tabular*}
\caption{Statistics of the datasets used in our experiments.}
\label{tab:dataset_statistics}
\vspace{-0.4cm}
\end{table}

\subsubsection{Baselines}
We compare \textsc{WhisperRec} with three groups of baselines: (1) ID-based sequential recommenders, including GRU4Rec~\cite{hidasi2015session}, BERT4Rec~\cite{sun2019bert4rec}, and HSTU~\cite{zhai2024actions}; (2) SID-based generative recommenders, including TIGER~\cite{rajput2023recommender} and OneRec~\cite{zhou2025onerec}; and (3) OneReason-based variants, including OneReason and OneReason-CoT \textit{No-Think} and \textit{Think} versions, serve as state-of-the-art FRM baselines. For CoT analysis, we compare MV-ACoT with standard OneReason-CoT and Adaptive CoT.

\subsubsection{Metrics and Implement}
For recommendation performance, we report SID Hit@K and ID Hit@K at $K=16,64$. For SID-based models, we map generated SIDs back to item IDs through SID decoding. For latent reasoning analysis, we report codebook retrieval Hit@1/2/3, inference QPS, semantic similarity, and MMLU scores. We implement \textsc{WhisperRec} with the Hugging Face \texttt{Trainer} and Fully Sharded Data Parallel (FSDP) for distributed training. For evaluation, we employ vLLM with beam search for efficient decoding. More implementation details are provided in Appendix A.

\subsection{Overall Results (RQ1)}
Table \ref{tab:WhisperRec_benchmark_compact} reports the overall recommendation performance on both the public benchmark and the  industrial dataset. \textsc{WhisperRec} denotes our latent reasoning model trained with MV-ACoT supervision.
Experimental results show that \textsc{WhisperRec} achieves the best performance across all metrics, outperforming traditional baselines and the state-of-the-art FRM OneReason. Compared with the explicit CoT Think and No-Think variants of OneReason, WhisperRec improves SID@64 by 17.44\% and 9.33\%, respectively. Its gains over OneReason and its explicit-CoT variants suggest that latent token reasoning is more effective than relying directly on verbose CoT supervision. Latent tokens may capture higher-level semantic reasoning patterns that better connect user behavior evidence, contextual intent, and target generation.

\begin{table}[t]
\centering
\scriptsize
\setlength{\tabcolsep}{3.6pt}
\renewcommand{\arraystretch}{0.88}
\begin{tabular*}{\columnwidth}{@{\extracolsep{\fill}}lccccc}
\toprule
\textbf{Dimension} & \textsc{OR} & \textsc{Ada} & \textsc{Exploration} & \textsc{Evaluation} & \textsc{Attribution} \\
\midrule
Factuality       & 3.48 & 3.70 & 3.96 & \textbf{4.04} & 3.62 \\
Evidence         & 3.17 & 3.68 & 4.00 & \textbf{4.34} & 4.13 \\
Signal           & 3.19 & 3.35 & 3.74 & \textbf{4.04} & 3.52 \\
Intent           & 3.76 & 3.86 & 4.52 & 4.59 & \textbf{4.68} \\
Causal           & 2.74 & 3.27 & 4.02 & \textbf{4.25} & 3.67 \\
Ad--User Fit     & 3.26 & 3.85 & 4.49 & \textbf{4.58} & 4.30 \\
Spatiotemporal   & 3.34 & 3.87 & 4.32 & 4.44 & \textbf{4.64} \\
Confidence       & 2.78 & 2.91 & 3.91 & \textbf{4.11} & 3.64 \\
Task Fulfillment & 2.56 & 3.36 & 4.48 & \textbf{4.59} & 4.08 \\
\midrule
\textbf{Average} & 3.14 & 3.54 & 4.16 & \textbf{4.35} & 4.03 \\
\bottomrule
\end{tabular*}
\caption{LLM-as-Judge results of CoT strategies.}
\label{tab:cot_quality_eval}
\vspace{-0.5cm}
\end{table}

\subsection{CoT Analysis (RQ2)}

We analyze our MV-ACoT from two aspects, the quality of teacher-generated CoT traces and their downstream effect on recommendation performance.

\subsubsection{LLM-as-a-Judge Evaluation}
We evaluate generated reasoning traces using an LLM-as-a-Judge protocol. For each CoT construction view, we sample 100 instances and score them with a unified rubric covering nine dimensions: \textbf{factuality}, {evidence selection}, \textbf{signal strength}, \textbf{intent translation}, \textbf{causal logic}, \textbf{ad-user fit}, \textbf{spatiotemporal reasoning}, \textbf{confidence}, and \textbf{task fulfillment}. We use GPT-5.5 as the judge; the complete prompt and rubric are provided in Appendix D.

As shown in Table \ref{tab:cot_quality_eval}. The results support the value of multi-view CoT construction. Although \textsc{Exploration} does not always achieve the highest overall score, it produces diverse user intent hypotheses that target-conditioned tasks may overlook. \textsc{Evaluation} provides stable candidate-level analysis by assessing whether an item is supported by current user evidence. \textsc{Attribution} reveals association chains behind observed conversions, particularly improving intent translation and spatiotemporal reasoning. Together, these views expand the intent space, improve evidence-grounded discrimination, and strengthen conversion-oriented reasoning, yielding richer supervision than a single CoT task.

\subsubsection{Ablation Study on CoT Effectiveness}

\begin{table}[t]
\centering
\scriptsize
\setlength{\tabcolsep}{3.6pt}
\renewcommand{\arraystretch}{0.92}
\begin{tabular*}{\columnwidth}{@{\extracolsep{\fill}}lcccc}
\toprule
\multirow{2}{*}{\textbf{CoT Construction}}
& \multicolumn{2}{c}{\textbf{Public Benchmark}}
& \multicolumn{2}{c}{\textbf{Industrial Dataset}} \\
\cmidrule(lr){2-3}
\cmidrule(l){4-5}
& \textbf{SID@64} & \textbf{ID@64}
& \textbf{SID@64} & \textbf{ID@64} \\
\midrule
\textsc{OneReason-CoT}
& 0.0092 & 0.0282
& 0.1939 & 0.5061 \\

\textsc{Adaptive CoT}
& 0.0144 & 0.0429
& 0.1935 & 0.5052 \\

\textsc{Exploration}
& 0.0144 & 0.0402
& 0.1971 & 0.5060 \\

\textsc{Evaluation}
& 0.0233 & 0.0712
& 0.1989 & 0.5070 \\

\textsc{Attribution}
& 0.0167 & 0.0560
& 0.1936 & 0.5055 \\

\textsc{MV-ACoT--Merge}
& \textbf{0.0239} & \textbf{0.0742}
& \textbf{0.1993} & \textbf{0.5088} \\
\bottomrule
\end{tabular*}
\caption{Ablation of CoT construction in \textsc{WhisperRec}.}
\label{tab:cot_ablation}
\vspace{-0.5cm}
\end{table}

We further examine whether MV-ACoT's complementary supervision translates into downstream gains for \textsc{WhisperRec}. As shown in Table~\ref{tab:cot_ablation}, replacing OneReason-CoT with Adaptive CoT already yields clear improvements, indicating the importance of matching reasoning complexity to recommendation difficulty. By avoiding unnecessary long-context reasoning for simple samples and focusing analysis on ambiguous cases, Adaptive CoT provides cleaner supervision and reduces the risk of hallucinated rationales.

\begin{figure}[h]
\vspace{-0.3cm}

\centering
\includegraphics[width=\columnwidth]{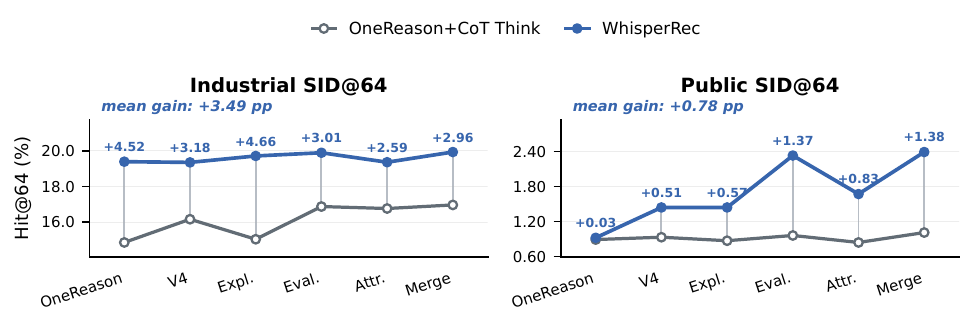}
\vspace{-0.9cm}
\caption{SID@64 performance comparison between OneReason Think and \textsc{WhisperRec} under different CoT.}
\vspace{-0.5cm}
\label{fig:sid_trend_mixed_vs_latent}
\end{figure}

Single-view MV-ACoT variants further improve performance through task-specific reasoning objectives. \textsc{Evaluation} performs strongly because candidate-level discrimination directly aligns with user--item decision making. \textsc{Exploration} broadens potential intent hypotheses, whereas \textsc{Attribution} captures conversion-oriented evidence chains. These results show that each view provides distinct reasoning signals beyond generic or adaptive CoT.
The integrated \textsc{MV-ACoT-Merge} setting further demonstrates the benefit of multi-view supervision. By combining intent exploration, evidence-grounded evaluation, and conversion attribution, it offers more comprehensive supervision than any single view. Figure~\ref{fig:sid_trend_mixed_vs_latent} also shows that \textsc{WhisperRec} consistently outperforms OneReason+CoT Think under the same CoT construction strategies, suggesting that latent reasoning more effectively translates view-specific supervision into SID retrieval gains. Improvements across Adaptive CoT and MV-ACoT variants further indicate that complexity adaptation and multi-view objectives complement latent encoding.

\subsection{Latent Reasoning Analysis (RQ3)}
In this section, we examine whether latent reasoning can further improve over explicit CoT reasoning.

\subsubsection{Recommendation Performance}
We first compare \textsc{UnCoT}, \textsc{CoT}, and \textsc{WhisperRec} on the recommendation task. As shown in Table~\ref{tab:latent_reasoning_effectiveness}, \textsc{CoT} yields limited and inconsistent gains over \textsc{UnCoT}, whereas latent reasoning substantially improves performance across all SID levels. In particular, \textsc{Latent Reason} improves Hit@L1 from $0.354$ to $0.401$ over explicit CoT and achieves the best results on Hit@L2 and Hit@L3. These results suggest that latent tokens more effectively encode reasoning information for recommendation.

\begin{table}[h]
\centering
\vspace{-0.2cm}
\scriptsize
\renewcommand{\arraystretch}{1.05}
\begin{tabular*}{\columnwidth}{@{\extracolsep{\fill}}lccc@{}}
\toprule
\textbf{Method} & \textbf{Hit@L1} & \textbf{Hit@L2} & \textbf{Hit@L3} \\
\midrule
\textsc{OneReason-UnCoT} & 0.350 & 0.122 & 0.045 \\
\textsc{OneReason-CoT} & 0.354 & 0.124 & 0.043 \\
\textsc{WhisperRec}
& \textbf{0.401} & \textbf{0.143} & \textbf{0.071} \\
\bottomrule
\end{tabular*}
\caption{Effectiveness of latent reasoning on SID prediction.}
\label{tab:latent_reasoning_effectiveness}
\vspace{-0.4cm}
\end{table}

\subsubsection{Inference Efficiency}
We compare the inference efficiency of \textsc{UnCoT}, \textsc{CoT}, and \textsc{WhisperRec} on 872 test instances. As shown in Table~\ref{tab:inference_efficiency}, \textsc{Plain CoT} requires autoregressive generation of rationales, substantially increasing latency and reducing throughput. In contrast, \textsc{WhisperRec} achieves latency and throughput nearly identical to \textsc{UnCoT}, while delivering a $17.49\times$ QPS gain over \textsc{Plain CoT}. These results show that latent reasoning avoids the autoregressive serving overhead of explicit CoT generation, making it highly suited for online recommendation systems.
\begin{table}[h]
\centering
\vspace{-0.2cm}
\scriptsize
\renewcommand{\arraystretch}{1.05}
\begin{tabular*}{\columnwidth}{@{\extracolsep{\fill}}lccc@{}}
\toprule
\textbf{Method} & \textbf{Time (s)} $\downarrow$ & \textbf{QPS} $\uparrow$ & \textbf{Speedup} $\uparrow$ \\
\midrule
\textsc{OneReason-UnCoT} & {3.17} & {274.56} & $17.51\times$ \\
\textsc{OneReason-CoT} & 55.59 & 15.68 & $1.00\times$ \\
\textsc{WhisperRec} & 3.17 & 274.29 & \textbf{17.49$\times$} \\
\bottomrule
\end{tabular*}
\vspace{-0.2cm}
\caption{Inference efficiency comparison. Speedup is measured relative to explicit CoT.}
\label{tab:inference_efficiency}
\vspace{-0.4cm}
\end{table}

\subsubsection{Interpretability}
We evaluate whether latent tokens retain decision-relevant semantics from explicit reasoning. For each MV-ACoT view, we reconstruct reasoning semantics from latent tokens and measure their cosine similarity to annotated Plain CoT counterparts using Qwen3-Emb-8B. 

As shown in Table~\ref{tab:semantic_similarity}, the reconstructed latent semantics achieve consistently high similarity across \textsc{Exploration}, \textsc{Evaluation}, and \textsc{Attribution} ($0.788$--$0.804$). These results provide evidence that latent tokens retain CoT semantics while reasoning in the latent space. Together with the retrieval and efficiency results, this analysis answers RQ3: latent reasoning improves recommendation performance and retains semantically aligned reasoning information, while achieving inference efficiency close to \textsc{UnCoT} models. Case study analyses are presented in Appendix~B.

\begin{table}[t]
\centering
\scriptsize
\renewcommand{\arraystretch}{1.12}
\begin{tabular*}{\columnwidth}{@{\extracolsep{\fill}}lccc@{}}
\toprule
\textbf{Metric} & \textsc{Exploration} & \textsc{Evaluation} & \textsc{Attribution} \\
\midrule
Cosine Similarity & 0.8027 & \textbf{0.8043} & 0.7878 \\
\bottomrule
\end{tabular*}
\caption{Semantic similarity reconstructed and explicit CoT.}
\label{tab:semantic_similarity}
\vspace{-0.7cm}
\end{table}

\subsubsection{Number of Latent Tokens}
\begin{figure}[h]
\centering
\vspace{-0.4cm}
\includegraphics[width=\columnwidth]{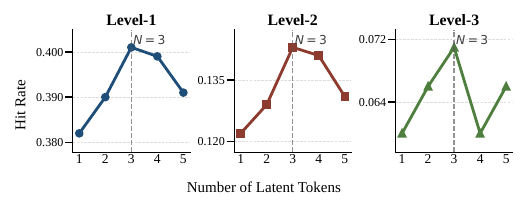}
\vspace{-0.7cm}
\caption{Ablation study on the number of latent tokens.}
\label{fig:latent_token_num_ablation}
\vspace{-0.6cm}
\end{figure}

We analyze the effect of the number of latent tokens on recommendation performance. As shown in Figure~\ref{fig:latent_token_num_ablation}, Hit improves as the number of tokens increases from one to three, indicating that multiple tokens help represent fine-grained reasoning information. Further increasing the number yields no consistent gains, likely due to redundancy. We therefore use three latent tokens by default to balance reasoning capacity and compactness.

\subsection{General Capability Analysis (RQ4)}

\begin{figure}[h]
\centering
\vspace{-0.4cm}

\includegraphics[width=0.5\columnwidth]{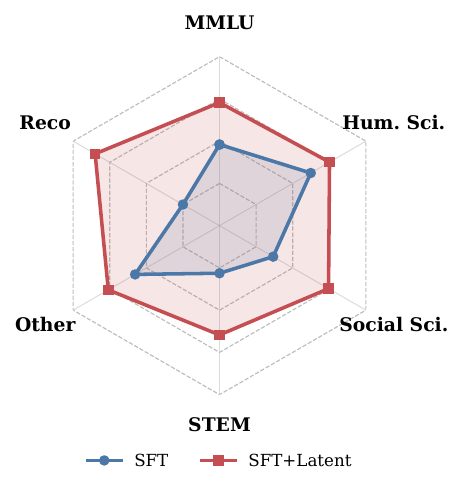}
\vspace{-0.4cm}

\caption{General and recommendation capability.}
\vspace{-0.6cm}

\label{fig:latent_radar}
\end{figure}

We further examine whether latent reasoning preserves general model capabilities. Figure \ref{fig:latent_radar} compares explicit CoT training and latent reasoning training on MMLU and recommendation quality scores. Latent reasoning achieves higher scores across all MMLU categories and also improves the recommendation score, indicating that latent supervision causes less degradation to general abilities than direct explicit CoT training while still strengthening recommendation reasoning.

\section{Conclusion}
In this paper, we propose \textsc{WhisperRec}, a latent reasoning framework for foundation recommendation models. Efficient reasoning is essential for deploying reasoning-enabled FRMs in industrial recommendation systems. Built upon \textsc{OneReason}, a strong FRM baseline, \textsc{WhisperRec} addresses the limitations of explicit CoT through MV-ACoT for diverse CoT construction, three-stage Latent Reasoning Alignment for internalizing teacher CoT into latent tokens, and curriculum-based post-training for downstream recommendation. Experiments on both public and industrial datasets show that \textsc{WhisperRec} outperforms conventional baselines and the \textit{Think} and \textit{No-Think} variants of \textsc{OneReason}. By reasoning in the latent space, \textsc{WhisperRec} achieves over $10\times$ higher throughput than explicit-CoT methods, demonstrating its potential for large-scale online recommendation.

\bibliography{ref}

\clearpage

\appendix
\section{Appendix A: Dataset Construction and Parameters Setting}
\label{app:parameters settings}

\subsection{Implementation details}
In our experiment, we perform full-parameter supervised fine-tuning on two complementary training streams, consisting of standard recommendation instances and chain-of-thought(CoT) augmented instances. Training is distributed over seven nodes with eight GPUs per node. We use a per-device batch size of 4 and one gradient accumulation step, resulting in an effective global batch size of 224. The model is optimized using AdamW with a peak learning rate of \(1.0\times10^{-4}\), a cosine learning-rate schedule, a warmup ratio of \(0.01\), and a weight decay of \(0.01\). We train the model for two effective epochs in BF16 precision and enable gradient checkpointing to reduce memory consumption. The maximum training sequence length is set to 4,096 tokens. For the streaming dataset, the number of optimization steps is dynamically determined from the total number of samples in the two training streams:

\begin{equation}
N_{\mathrm{step}}
=
\left\lceil
\frac{
    (N_{1}+N_{2})E
}{
    N_{\mathrm{node}}
    N_{\mathrm{GPU}}
    B_{\mathrm{device}}
    N_{\mathrm{acc}}
}
\right\rceil,
\end{equation}
where \(N_{1}\) and \(N_{2}\) denote the numbers of samples in the two streams, \(E=2\) is the target number of epochs, \(N_{\mathrm{node}}=7\), \(N_{\mathrm{GPU}}=8\), \(B_{\mathrm{device}}=4\), and \(N_{\mathrm{acc}}=1\). We use a fixed random seed of 42 for all experiments.

\begin{table}[h]
\centering

\small
\renewcommand{\arraystretch}{1.08}
\setlength{\tabcolsep}{3pt}
\begin{tabular}{
@{}
>{\raggedright\arraybackslash}p{0.19\linewidth}
>{\raggedright\arraybackslash}p{0.45\linewidth}
>{\raggedright\arraybackslash}p{0.30\linewidth}
@{}
}
\toprule
\textbf{Category} & \textbf{Hyperparameter} & \textbf{Value} \\
\midrule

\multirow{4}{*}{\makecell[l]{Model\\configuration}}
& Initialization & Qwen3.5-0.8B \\
& Fine-tuning strategy & Full-parameter fine-tuning \\
& Numerical precision & BF16 \\
& Maximum sequence length & 4,096 tokens \\
\midrule

\multirow{2}{*}{\makecell[l]{Semantic\\IDs}}
& Number of semantic levels & 3 \\
& Codebook size per level & 1,024 \\
\midrule

\multirow{6}{*}{Optimization}
& Optimizer & AdamW \\
& Learning rate & $1.0 \times 10^{-4}$ \\
& Scheduler & Cosine \\
& Warmup ratio & 0.01 \\
& Weight decay & 0.01 \\
& Gradient checkpointing & Enabled \\
\midrule

\multirow{3}{*}{Training}
& Epochs & 2 \\
& Gradient accumulation steps & 1 \\
& Random seed & 42 \\
\midrule

\multirow{2}{*}{\makecell[l]{Streaming\\data}}
& Number of training streams & 2 \\
& Maximum training steps & $\left\lceil 2(N_1 + N_2) / 224 \right\rceil$ \\
\midrule

\multirow{3}{*}{Checkpoint}
& Logging interval & 10 steps \\
& Checkpoint interval & 500 steps \\
& Maximum retained checkpoints & 5 \\
\midrule

\multirow{2}{*}{Hardware}
& Number of nodes & 7 \\
& GPUs per node & 8 \\
\bottomrule
\end{tabular}

\caption{Hyperparameters for multi-stream SFT}
\label{tab:multi_stream_hyperparameters}
\end{table}

\subsection{Construction of the supervised fine-tuning dataset.}
We construct the supervised fine-tuning dataset from temporally ordered user-item interaction logs. Each item is associated with a hierarchical semantic identifier (SID), while each user is associated with a compact profile containing general preferences, local-life demands, and, when available, explicitly expressed intents.

For each user, we first sort all interactions chronologically and remove redundant interactions with the same item. When multiple interaction signals are associated with a duplicated item, we retain the strongest feedback according to the following priority:
\[
\text{conversion} >
\text{click} >
\text{long view} >
\text{impression}.
\]
Each conversion event is subsequently treated as a prediction target. Specifically, for the \(k\)-th conversion event, all interactions occurring before it form the historical context, whereas the SID of the converted item serves as the supervision signal. In this way, a user with multiple conversion events may contribute multiple training instances without including the target item in its own input history.

To control the context length and emphasize recent interests, we retain at most the latest 50 interactions before each target event. Historical items are represented by their hierarchical SIDs and augmented with their corresponding behavior types, allowing the model to distinguish weak feedback, such as impressions, from stronger preference signals, such as clicks and conversions. The resulting input combines the compact user profile, local-life demand, explicit intent, and behavior-aware SID sequence. The output is the hierarchical SID of the next converted item. Formally, each training instance is written as
\begin{equation}
\mathcal{H}_{u,k}
=
\bigl[(s_i, a_i)\bigr]_{i=\max(1,\,k-L)}^{k-1},
\end{equation}

\begin{equation}
\mathcal{X}_{u,k}
=
[P_u;\, D_u;\, I_u;\, \mathcal{H}_{u,k}],
\qquad
\mathcal{Y}_{u,k} = s_k.
\end{equation}
where \(P_u\), \(D_u\), and \(I_u\) denote the user profile, user demand, and explicit intent, respectively; \(s_i\) and \(a_i\) denote the SID and behavior type of the \(i\)-th interaction; and \(L=50\) is the maximum history length.

We discard instances whose user profile or target SID is unavailable. Historical items without valid SID mappings are filtered from the input sequence. Finally, all valid samples are stored in sharded JSONL files to support memory-efficient streaming and distributed training.

\subsection{Construction of CoT supervision.}
We employ Qwen3-235B-A22B-Instruct-2507\footnote{\url{https://huggingface.co/Qwen/Qwen3-235B-A22B-Instruct-2507}} as the teacher model to generate CoT supervision. The teacher takes the user profile, local-life demand, explicit intent, and historical interactions as input. The subsequent converted item is provided only as privileged guidance, and the teacher is instructed not to reveal it directly. The resulting concise rationale is combined with the ground-truth SID to construct the CoT-enhanced supervised fine-tuning data.

\begin{table}[t]
\centering
\small

\renewcommand{\arraystretch}{1.08}
\setlength{\tabcolsep}{3pt}

\begin{tabular}{
@{}
>{\raggedright\arraybackslash}p{0.25\linewidth}
>{\raggedright\arraybackslash}p{0.67\linewidth}
@{}
}
\toprule
\textbf{Component} & \textbf{Configuration} \\
\midrule

Teacher model
& Qwen3-235B-A22B-Instruct-2507 \\
\midrule

User information
& User profile, user demand, and explicitly expressed intent \\
\midrule

Interaction history
& Item descriptions, behavior types, timestamps, and geographic context \\
\midrule

Privileged guidance
& Description of the subsequent converted item, used only to guide reasoning without directly revealing the target \\
\midrule

Generated supervision
& A concise rationale connecting historical behavior with the user's potential commercial intent \\
\midrule

Final training target
& Teacher-generated rationale combined with the ground-truth SID \\
\bottomrule
\end{tabular}

\caption{Configuration for constructing CoT supervision.}
\label{tab:cot_construction}
\end{table}

\section{Appendix B: Case Study of Latent Reconstruction}
\label{sec:qualitative_reconstruction}

\begin{figure*}[tp]
  \centering
  \includegraphics[
    width=\textwidth,
    trim=100 100 60 100,
    clip
  ]{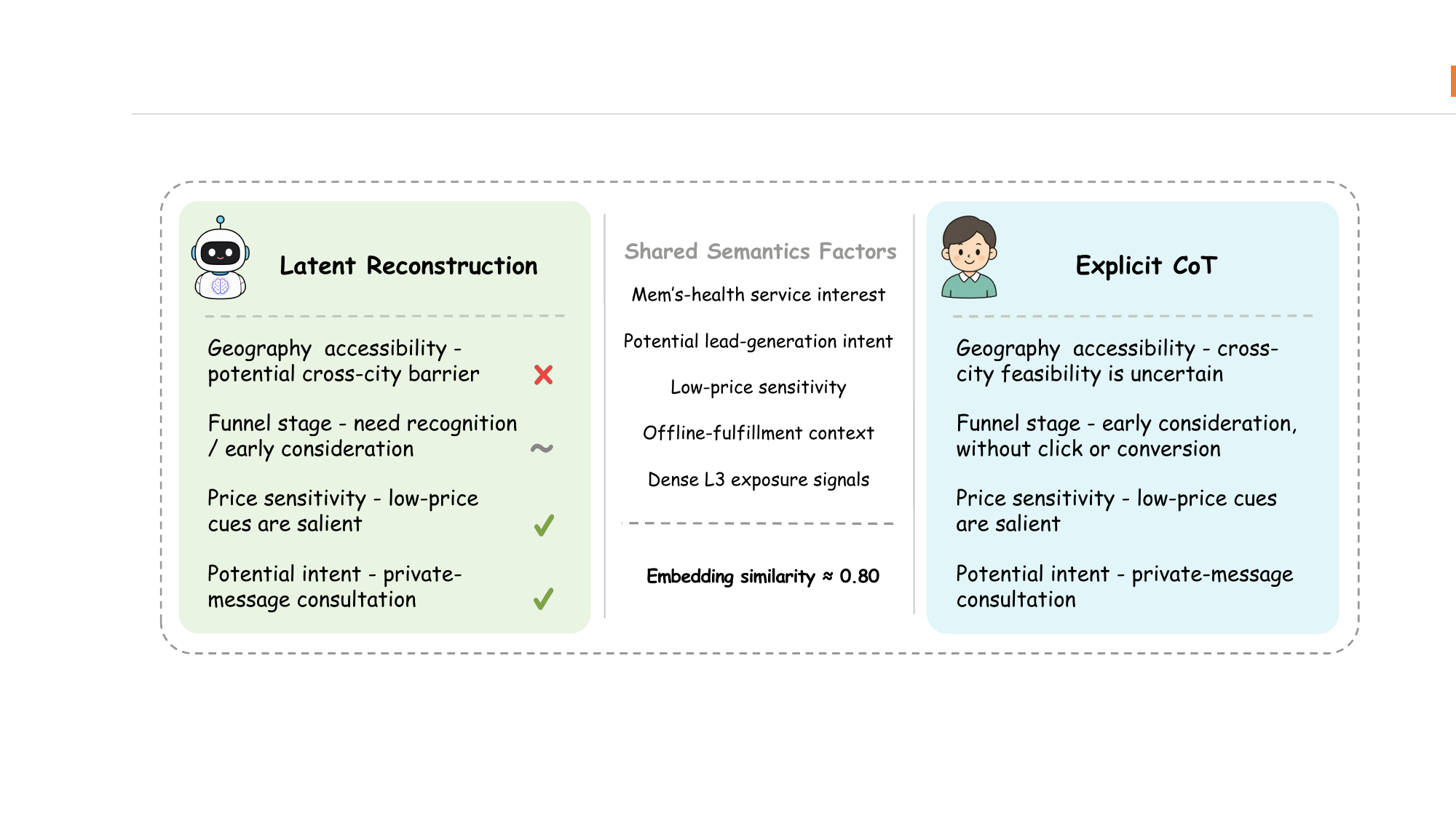}
  \caption{Comparison between latent reconstruction and explicit CoT. The two traces share core commercial semantics, including price sensitivity, consultation intent, and funnel-stage judgment, but differ in geographic accessibility. Symbols denote aligned ($\checkmark$), partially aligned ($\sim$), and divergent ($\times$) factors.}
  \label{fig:semantic_reconstruction}
\end{figure*}

Figure~\ref{fig:semantic_reconstruction} presents a qualitative comparison between latent reconstruction and explicit CoT. Despite operating in a continuous latent space, the reconstructed trace preserves the main commercial semantics in the explicit rationale, including men's-health interest, potential lead-generation intent, price sensitivity, offline fulfillment, and dense L3 behavioral signals. The two traces are largely aligned on price sensitivity, consultation intent, and conversion-funnel stage, while differing in their assessment of geographic accessibility.

With an embedding similarity of approximately $0.80$, this case provides qualitative evidence that latent tokens retain core decision-relevant semantics without directly reproducing the surface form of explicit CoT. The remaining divergence also suggests that latent reasoning can encode semantically aligned but non-identical decision paths.

\section{Appendix C: MV-ACoT Prompt Construction Principles}

\label{app:mv_acot_prompt}
Beyond the complementary Exploration, Evaluation, and Attribution tasks, MV-ACoT introduces business-aware prompt constraints to improve the quality and reliability of teacher-generated reasoning. As summarized in Table~\ref{tab:mv_acot_prompt_principles}, these constraints ground the teacher model in local-service lead-ad recommendation, requiring it to translate content interests into actionable commercial demands and prioritize strong behavioral evidence over weak exposures. This prevents the teacher from relying on generic content preference descriptions that are insufficient for conversion-oriented recommendation decisions.

MV-ACoT further adapts reasoning paths to the strength, consistency, and uncertainty of user signals. It incorporates fulfillment mode, geographic constraints, industry-specific factors, and conversion stage when analyzing user intent. Moreover, evidence references and low-confidence or no-attribution outputs discourage hallucination, target leakage, and forced explanations. Together, these constraints enable concise reasoning for clear cases while supporting sufficiently detailed, evidence-grounded analysis for challenging recommendation scenarios.

\begin{table*}[tp]
\centering

\renewcommand{\arraystretch}{1.15}
\begin{tabular}{p{0.20\textwidth} p{0.38\textwidth} p{0.34\textwidth}}
\toprule
\textbf{Prompt Principle} & \textbf{Design} & \textbf{Effect} \\
\midrule
Role and task specification
& Define the teacher model as a local-service and lead-ad recommendation expert, with the task restricted to lead-ad interaction and conversion reasoning.
& Aligns the reasoning objective with the advertising recommendation scenario and avoids drifting to generic content recommendation or item description. \\

Interest-to-demand translation
& Require the model to map content interests into commercial needs that can be served by lead advertisements.
& Upgrades CoT from describing ``what the user likes'' to inferring ``what commercial demand the user may have.'' \\

Behavior signal hierarchy
& Categorize user behaviors into L1 conversion-level, L2 click-level, and L3 exposure-level signals, and prioritize evidence strength over frequency.
& Prevents shallow exposures from being mistaken as strong intent and improves key evidence selection. \\

Adaptive reasoning path selection
& Select fast convergence, competitive comparison, deep reasoning, or low-cost light-conversion reasoning according to the signal structure.
& Enables concise reasoning for easy cases and multi-chain comparison for difficult cases, reducing redundant reasoning. \\

Fulfillment and industry knowledge
& Distinguish online and offline fulfillment, and incorporate location, service radius, industry decision factors, and conversion stage.
& Improves adaptation to local-service advertising, spatiotemporal constraints, and industry-specific conversion logic. \\

Evidence grounding and anti-hallucination
& Require references to concrete item IDs and allow low-confidence, weak-attribution, or no-attribution conclusions.
& Reduces fabricated behaviors, target leakage, and forced explanations, strengthening evidence grounding. \\
\bottomrule
\end{tabular}

\caption{Prompt construction principles of MV-ACoT for local-service lead-ad recommendation. Beyond the multi-view task design, the prompts incorporate domain-specific role constraints, demand-oriented intent translation, hierarchical behavior signals, adaptive reasoning paths, fulfillment-aware industry knowledge, and evidence-grounding mechanisms.}
\label{tab:mv_acot_prompt_principles}
\end{table*}

\section{Appendix D: CoT Quality Evaluation Prompt}
\label{app:cot_quality_prompt}

We employ an LLM-as-a-Judge protocol to evaluate the quality of teacher-generated reasoning traces for local-service lead-ad recommendation. Rather than assessing whether a trace simply supports a candidate or conversion target, the protocol evaluates whether it derives a calibrated recommendation judgment from available user evidence. For each instance, the judge receives the task type, user history, supplementary user profile, candidate or converted advertisement, and generated reasoning trace.

User history is treated as the primary factual evidence. User-profile attributes and target information may provide context, but must not be backfilled into historical behaviors or used to fabricate supporting evidence. As summarized in Tables~\ref{tab:judge_task_principles} and~\ref{tab:judge_dimensions}, the protocol combines task-specific requirements with a unified multi-dimensional scoring rubric. This design evaluates not only factuality and evidence selection, but also intent translation, causal logic, advertising fit, spatiotemporal reasoning, confidence calibration, and task fulfillment.

\paragraph{Hard Scoring Constraints.}
To enforce factual faithfulness, the judge applies explicit score caps. If a trace claims clicks, conversions, searches, consultations, or lead submissions absent from the user history, its factuality score cannot exceed 2. Similarly, backfilling candidate or target information into the user history also limits factuality to at most 2. Fabricated locations, behavior times, consumption stages, or explicit user needs limit the score to at most 3, with lower scores assigned to severe violations. These constraints prioritize evidence-grounded reasoning over target consistency or explanatory completeness.

\paragraph{Output Format.}
The judge returns a single JSON object containing the nine dimensions in Table~\ref{tab:judge_dimensions}. Each dimension includes an integer \texttt{score} from 1 to 5 and a brief \texttt{reason}. No markdown, additional fields, or invalid score values are permitted. Each evaluation instance is formatted as follows:

\colorbox{black!7}{%
\begin{minipage}{0.94\linewidth}

\small
\textbf{Evaluation Instance Format.}

\begin{itemize}[leftmargin=1.25em,itemsep=1pt,topsep=2pt]
    \item \textbf{Task type:} \texttt{\{\{task\_type\}\}}
    
    \item \textbf{Historical behaviors:} \texttt{\{\{history\_text\}\}}
    
    \item \textbf{User profile:} \texttt{\{\{user\_plain\_text\}\}}
    
    \item \textbf{Candidate or target item:} \texttt{\{\{target\_text\}\}}
    
    \item \textbf{Generated reasoning trace:} \texttt{\{\{reasoning\_text\}\}}
\end{itemize}
\end{minipage}%
}

\begin{table*}[p]
\centering
\vspace*{\fill}

\setlength{\tabcolsep}{2pt}
\renewcommand{\arraystretch}{1.38}
\setlength{\extrarowheight}{1.5pt}
\begin{tabular*}{\textwidth}{@{\extracolsep{\fill}}
    p{0.15\textwidth}
    p{0.32\textwidth}
    p{0.45\textwidth}@{}}
\toprule
\textbf{Component} & \textbf{Definition} & \textbf{Evaluation Focus} \\
\midrule
\multicolumn{3}{@{}l}{\textit{Task-specific requirements}} \\
\addlinespace[2pt]

Exploration
& Infer multiple potential commercial intents from user profiles and historical behaviors without observing a target advertisement.
& Assess whether the trace identifies two or three plausible, diverse intent directions, translates interests into serviceable demands, and provides calibrated confidence or priority rankings. \\

Evaluation
& Assess a candidate advertisement without assuming that it is a correct recommendation.
& Assess whether the candidate is supported or rejected based on historical evidence, signal strength, user demand, and contextual constraints. Negative or low-confidence judgments are allowed. \\

Attribution
& Explain an observed conversion to a target advertisement.
& Assess whether historical evidence supports the conversion. When evidence is insufficient, weak or no attribution is preferred to post-hoc rationalization. \\

\midrule
\multicolumn{3}{@{}l}{\textit{General evaluation principles}} \\
\addlinespace[2pt]

Evidence priority
& Treat \texttt{history\_text} as the primary factual source and \texttt{user\_plain\_text} as supplementary information.
& Profile attributes and target information must not override, alter, or be presented as historical behavioral facts. \\

Behavior hierarchy
& Rank signals by conversion depth: conversion $>$ click $>$ exposure or viewing for more than three seconds.
& Repeated shallow exposures should not be treated as stronger intent than fewer clicks or conversions. \\

Fulfillment awareness
& Distinguish between online and offline fulfillment.
& Offline services should consider location, service radius, accessibility, and demand timing; online services need not include geographic analysis unless a clear conflict exists. \\

Uncertainty awareness
& Permit mismatch, low-confidence, weak-attribution, and no-attribution conclusions.
& Reward factual faithfulness and calibrated reasoning rather than whether a trace supports the candidate or target. \\

Anti-hallucination
& Prohibit fabricated clicks, conversions, searches, consultations, lead submissions, locations, behavior times, or explicit needs.
& Prioritize factuality over target consistency or explanatory completeness; traces must not invent evidence to support a target. \\
\bottomrule
\end{tabular*}

\caption{Task-specific requirements and general principles for LLM-based CoT quality evaluation.}
\label{tab:judge_task_principles}
\vspace*{\fill}
\end{table*}

\begin{table*}[p]
\centering
\vspace*{\fill}

\setlength{\tabcolsep}{2pt}
\renewcommand{\arraystretch}{1.24}
\setlength{\extrarowheight}{1.2pt}
\begin{tabular*}{\textwidth}{@{\extracolsep{\fill}}
    p{0.14\textwidth}
    p{0.26\textwidth}
    p{0.18\textwidth}
    p{0.18\textwidth}
    p{0.18\textwidth}@{}}
\toprule
\textbf{Dimension}
& \textbf{Evaluation Criterion}
& \multicolumn{3}{c}{\textbf{Scoring Anchors}} \\
\cmidrule(l){3-5}
& & \textbf{Low (1)} & \textbf{Medium (3)} & \textbf{High (5)} \\
\midrule

Factuality
& Faithfulness to historical behaviors without fabrication, exaggeration, chronological confusion, or target backfilling.
& Substantial fabrication or severe detachment from history.
& Mostly based on history but contains local misreadings or overgeneralization.
& Fully grounded in history with no evident fabrication or backfilling. \\

Evidence selection
& Selection of the most relevant, strong, recent, and decision-relevant historical evidence.
& Relies on irrelevant evidence and ignores key signals.
& Uses partially relevant evidence but includes noise or misses competing signals.
& Accurately prioritizes key evidence and handles noise and competing directions. \\

Signal strength
& Correct weighting of conversions, clicks, and shallow exposures according to conversion depth.
& Treats weak exposure as decisive or ignores strong conversion evidence.
& Generally distinguishes signal levels but occasionally overweights frequency.
& Robustly weights behavior depth, direction concentration, and temporal order. \\

Intent translation
& Translation from content interests into local-service needs and actionable advertising intent.
& Remains entirely at the content-interest level.
& Identifies a service category but provides limited conversion context.
& Clearly identifies service needs, consumption scenarios, and potential conversion actions. \\

Causal logic
& Coherence of the chain from historical behavior to need, commercial intent, and final judgment.
& Contains severe logical gaps, contradiction, or forced causality.
& The main logic holds but some inference steps or stage transitions are weak.
& Forms a clear and well-supported reasoning chain with appropriate uncertainty. \\

Advertising fit
& Alignment with lead-ad recommendation rather than ordinary content recommendation.
& Only performs generic content matching.
& Identifies an advertising category but lacks conversion or decision factors.
& Characterizes service needs, conversion paths, and major commercial decision factors. \\

Spatiotemporal reasoning
& Appropriate handling of time, location, fulfillment mode, service radius, and accessibility.
& Ignores clear temporal or regional conflicts.
& Partially considers time or location but insufficiently distinguishes fulfillment modes.
& Applies restrained and appropriate spatiotemporal reasoning according to fulfillment type. \\

Confidence calibration
& Consistency between conclusion strength or confidence and the quality of supporting evidence.
& Is clearly overconfident, underconfident, or internally inconsistent.
& Confidence is generally reasonable but locally insufficiently calibrated.
& Confidence accurately reflects evidence strength, logical completeness, and task uncertainty. \\

Task fulfillment
& Completion of the requirements associated with Exploration, Evaluation, or Attribution.
& Fails to perform the requested reasoning task.
& Completes the main task but omits diversity, confidence, attribution, or a clear judgment.
& Fully follows the task-specific objective and required output structure. \\
\bottomrule
\end{tabular*}

\caption{Scoring dimensions used by the LLM judge. Each dimension is rated on a five-point scale; representative low-, medium-, and high-quality anchors are provided.}
\label{tab:judge_dimensions}
\vspace*{\fill}
\end{table*}

\end{document}